\documentclass[11pt]{article}
\usepackage{graphicx} 
\usepackage{amsmath}
\usepackage{amsfonts}   
\usepackage{amssymb}

\usepackage{setspace}
\begin{document}
\date{Today}
\title{{\bf{\Large Voros product, Noncommutative Schwarzschild Black Hole 
and Corrected Area Law }}}

\author{
{\bf {\normalsize Rabin Banerjee}$
$\thanks{E-mail: rabin@bose.res.in}}
\\ {\normalsize S.~N.~Bose National Centre for Basic Sciences,}
\\{\normalsize JD Block, Sector III, Salt Lake, Kolkata-700098, India}
\\[0.3cm]
{\bf {\normalsize Sunandan Gangopadhyay}$
$\thanks{E-mail: sunandan.gangopadhyay@gmail.com, sunandan@bose.res.in}}{\footnote{Also, Visiting Associate at S.~N.~Bose National Centre for Basic Sciences, JD Block, Sector III, Salt Lake, Kolkata-700098, India.}}\\
 {\normalsize West Bengal State University,}
\\{\normalsize Barasat, North 24 Paraganas, West Bengal, India}
\\[0.3cm]
{\bf {\normalsize Sujoy Kumar Modak}$
$\thanks{E-mail: sujoy@bose.res.in}}
\\ {\normalsize S.~N.~Bose National Centre for Basic Sciences,}
\\{\normalsize JD Block, Sector III, Salt Lake, Kolkata-700098, India}
\\[0.3cm] 
}
\date{}

\maketitle

\begin{abstract}
{\noindent We show the importance of the Voros product 
in defining a noncommutative Schwarzschild black hole. 
The corrected entropy/area-law is then computed in the tunneling
formalism. Two types of corrections are considered; one, due to the effects
of noncommutativity and the other, due to the effects of going beyond the
semiclassical approximation. The leading correction to the semiclassical
entropy/area-law is logarithmic and its 
coefficient involves the noncommutative
parameter.}
\end{abstract}
\vskip 1cm

\section{Introduction}

The theoretical discovery of radiation from black holes by Hawking
\cite{Hawking1, Hawking2} disclosed the first physically relevant window on the 
mysteries of quantum gravity. The analysis is based on 
quantum field theory in a curved background and reveals 
that black holes emit a spectrum that is similar to a thermal black body
spectrum. This result made the {\it first law of black hole mechanics} \cite{Bardeen} closely analogous to the first law of thermodynamics. This analogy ultimately led to the entropy for black holes and was consistent with the proposal made by Bekenstein \cite{Beken1,Beken2,Beken3,Beken4} that a black hole has an entropy proportional to its horizon area. All the above issues finally led to the famous Bekenstein-Hawking area law for the entropy of black holes given by $S_{BH}=A/4$. In a recent analysis \cite{Modakex} it was proved that, taking the  black hole entropy to be a {\it state function}, the standard Bekenstein-Hawking area law follows without using the {\it first law of black hole mechanics}. 

However, most of these calculations were based on a semiclassical
treatment and also on a commutative spacetime. The standard
Bekenstein-Hawking area law is known to get corrections due to
quantum geometry or back reaction effects \cite{list}.
Recently, modifications to the semiclassical 
area law due to noncommutative (NC) spacetime have also been 
obtained \cite{Modaknon}-\cite{Smail}. 
The motivation for these investigations 
was that noncommutativity is expected to be relevant at the 
Planck scale where it is known that usual semiclassical 
considerations break down. 
To obtain NC effects on the usual area law, 
computation of the (NC) Hawking temperature was done. 
The Gibbs form of first law of thermodynamics 
was finally used to obtain the entropy. 
It was observed \cite{Modaknon}, \cite{Banrev} that to the leading order in the NC parameter $\theta$, 
in the regime $\frac{r^2_h}{4\theta}>>1$, 
the (NC) area law was just a NC deformation 
of the usual semiclassical area law. 
Further, a graphical analysis revealed that 
when $r_h\geq4.8\sqrt\theta$, the NC version 
of the area law holds good to all orders in $\theta$. 

Inspite of this reasonable literature  \cite{Modaknon}-\cite{Smail}, \cite{lopez} on noncommutative black holes, two outstanding issues remain. First, there is no clear cut connection of this type of noncommutativity with our standard notions of a NC spacetime where point-wise multiplications are replaced by appropriate star multiplications. Secondly, computations are confined strictly within the semiclassical regime. The present analysis precisely addresses these issues.

In this paper, we have two objectives. First of all,
we shall point out that the Voros star product \cite{Voros,Lizzi} 
plays an important role in the obtention of the 
mass density of a static, spherically symmetric,
smeared, particle-like gravitational source. This observation is completely
new and has been missing in the existing literature 
\cite{Modaknon}-\cite{Smail}, \cite{lopez}. 
To uncover the significance of the Voros product in writing down
the mass density, it is important to take a look at the formulation
and interpretational aspects of NC quantum mechanics 
\cite{gouba},\cite{sunandan}. We observe that the inner product of the
coherent states (used in the construction of the wave-function 
of a ``free point particle"), from which the mass density is written down,
can be computed by using a deformed completeness relation 
(involving the Voros product) among the coherent
states. Further, one can give 
a consistent probabilistic interpretation of the wave-function 
of the ``free point particle" only when the Voros product is incorporated.
Indeed the Gaussian distribution which follows from this Voros type 
interpretation naturally includes the effect of the NC parameter $\theta$
and agrees with the structure given in the literature 
\cite{Modaknon}-\cite{Smail}, \cite{lopez}.

Our second objective is to find quantum corrections 
to the semiclassical Hawking temperature and entropy for 
noncommutative Schwarzschild black hole. 
This is done by first computing the correction 
to the Hawking temperature by going beyond the 
semiclassical approximation in the tunneling method 
\cite{Majhibeyond}-\cite{Tao}, \cite{Modakex}. 
Using the corrected form of the Hawking temperature 
and the first law of black hole thermodynamics, 
the entropy is computed. In the literature 
there are lot of papers which discuss the 
quantum corrections to above entities for 
different black holes in many different 
approaches \cite{Fursaev,Partha,Das,Carlip,Hooft,Modakex,Majhitrace}. 
However all of these works are confined to the 
commutative case only and there is nothing concerning  
the possible quantum corrections to the 
semiclassical Hawking temperature and entropy for 
noncommutative black holes. Here we want to 
generalise our method \cite{Modakex,Majhitrace} 
to the case of the NC Schwarzschild 
black hole and derive the appropriate corrections to the black hole entropy. 
The result is seen to contain logarithmic and inverse horizon area 
corrections and holds to $\mathcal{O}(\sqrt{\theta}e^{-M^2/\theta})$. 
The coefficient of the logarithmic correction term is explicitly determined 
from the trace anomaly of the stress tensor \cite{Hawking3},\cite{Dewitt}. 
This coefficient is also found to have 
NC correction and holds to the same order as the logarithmic correction 
to the entropy. 

The paper is organised as follows. In section 2, we shall
discuss about the role of the Voros product in defining a noncommutative
Schwarzschild black hole. In section 3, the corrections to the area law
by the tunneling method are discussed. In section 4, we present the
computation of the coefficient of the logarithmic correction to the area law.
Finally, we conclude in section 5.


\section{Voros product and noncomutative Schwarzschild black hole}

In this section, we present the role played by the Voros product
in defining the mass density of the NC Schwarzschid black hole.
To begin the discussion, we note that the NC
effect in gravity can be included in several ways
\cite{riv, wess1, wess2, lopez, okon, chaichian, saha, muk}. 
One of the ways in which the noncommutative effect
is incorporated is to take the spacetime as noncommutative
$[x_{\mu}, x_{\nu}]=i\theta_{\mu\nu}$ and use the Seiberg-Witten
map to recast the gravitational theory in noncommutative space in terms
of the corresponding theory in  usual commutative space leading to
correction terms involving powers of $\theta_{\mu\nu}$ in the various
expressions such as the metric, Riemann tensor, etc 
\cite{lopez}-\cite{muk}. This method naturally implies the use
of the Moyal star product \cite{szabo, sun1}. 
The twisted formulation of NC quantum
field theory \cite{chaichian1}-\cite{sun2} is yet another way 
of incorporating the effects of
noncommutativity in gravity \cite{wess1, wess2}. Alternatively,
one can incorporate the effect of noncommutativity in the mass term
of the gravitating object. This is done by representing the mass density  
by a Gaussian distribution instead of a 
Dirac delta distribution \cite{spal1},\cite{Smail}.
These ways of including noncommutative effects in gravity, are in general
inequivalent. Also, the specific nature of the NC product, if any, that
replaces the point-wise product, remains completely obscure.
In this paper, we shall employ the latter route 
to include the effect of noncommutativity in gravity. 
We shall show that there is a subtle connection
between the latter approach and the Voros star product just as there
is a direct connection between the former approaches and the Moyal star
product. To see this connection, we must first pay a careful attention to the
formulation and interpretational aspects of NC quantum mechanics. We
present a brief review of these issues for the sake of completeness
and also to pin down the importance of the Voros product in defining
the mass density of the NC Schwarzschild black hole.

Recently, in a couple of papers \cite{gouba},\cite{sunandan}, it was
suggested that NC quantum mechanics should be formulated as a quantum
system on the Hilbert space of Hilbert-Schmidt operators acting on classical
configuration space. In two dimensions, 
the coordinates of NC configuration 
space satisfy the commutation relation 
\begin{equation}
[\hat{x}, \hat{y}] = i\theta\quad,\quad\theta>0~,
\label{100a}
\end{equation} 
for a constant $\theta$. The annihilation and creation operators defined by
$b = \frac{1}{\sqrt{2\theta}} (\hat{x}+i\hat{y})$,
$b^\dagger =\frac{1}{\sqrt{2\theta}} (\hat{x}-i\hat{y})$
satisfy the Fock algebra $[ b ,b^\dagger ] = 1$. 
The NC configuration space is then 
isomorphic to the boson Fock space
\begin{eqnarray}
\mathcal{H}_c = \textrm{span}\{ |n\rangle= 
\frac{1}{\sqrt{n!}}(b^\dagger)^n |0\rangle\}_{n=0}^{n=\infty}
\label{300a}
\end{eqnarray}
where the span is taken over the field of complex numbers.
The next step is to introduce the Hilbert space
of the NC quantum system. We consider the set of 
Hilbert-Schmidt operators acting on NC configuration space
\begin{equation}
\mathcal{H}_q = \left\{ \psi(\hat{x},\hat{y}): 
\psi(\hat{x},\hat{y})\in \mathcal{B}
\left(\mathcal{H}_c\right),\;
{\rm tr_c}(\psi^\dagger(\hat{x},\hat{y})
\psi(\hat{x},\hat{y})) < \infty \right\}.
\label{400a}
\end{equation}
Here ${\rm tr_c}$ denotes the trace over NC 
configuration space and $\mathcal{B}\left(\mathcal{H}_c\right)$ 
the set of bounded operators on $\mathcal{H}_c$. 
To distinguish states in the NC configuration space from those in the
quantum Hilbert space, we denote states in the NC configuration 
space by $|\cdot\rangle$ and states in the 
quantum Hilbert space by $\psi(\hat{x},\hat{y})\equiv |\psi)$. 
Assuming commutative momenta, a unitary representation 
of the NC Heisenberg algebra in terms of operators 
$\hat{X}$, $\hat{Y}$, $\hat{P}_x$ and $\hat{P}_y$ 
acting on the states of the quantum Hilbert space 
(\ref{400a}) is easily found to be\footnote{We use capital letters to
distinguish operators acting on quantum Hilbert space from those acting
on NC configuration space.} 
\begin{eqnarray}
\label{schnc}
\hat{X}\psi(\hat{x},\hat{y}) &=& \hat{x}\psi(\hat{x},\hat{y})\quad,\quad
\hat{Y}\psi(\hat{x},\hat{y}) = \hat{y}\psi(\hat{x},\hat{y})\nonumber\\
\hat{P}_x\psi(\hat{x},\hat{y}) &=& \frac{\hbar}{\theta}
[\hat{y},\psi(\hat{x},\hat{y})]\quad,\quad
\hat{P}_y\psi(\hat{x},\hat{y}) = -\frac{\hbar}{\theta}
[\hat{x},\psi(\hat{x},\hat{y})]~.
\end{eqnarray}
The minimal uncertainty states on NC 
configuration space, which is isomorphic to boson Fock space, 
are well known to be the normalized coherent states \cite{klaud}
\begin{equation}
\label{cs} 
|z\rangle = e^{-z\bar{z}/2}e^{z b^{\dagger}} |0\rangle
\end{equation}
where, $z=\frac{1}{\sqrt{2\theta}}\left(x+iy\right)$ 
is a dimensionless complex number. These states provide an overcomplete 
basis on the NC configuration space. 
Corresponding to these states we can construct a state 
(operator) in quantum Hilbert space as follows
\begin{equation}
|z, \bar{z} )=\frac{1}{\sqrt{\theta}}|z\rangle\langle z|.
\label{csqh}
\end{equation}
These states have the property
\begin{equation}
B|z, \bar{z})=z|z, \bar{z})\quad,\quad B=\frac{1}{\sqrt{2\theta}}
(\hat{X}+i\hat{Y})\quad.
\label{p1}
\end{equation}
We now introduce the momentum eigenstates normalised such that
$(p'|p)=\delta(p'-p)$
\begin{eqnarray}
|p)=\sqrt{\frac{\theta}{2\pi\hbar^{2}}}e^{i\sqrt{\frac{\theta}{2\hbar^2}}
(\bar{p}b+pb^\dagger)}\quad,\quad\hat{P}_{i}|p)=p_{i}|p)
\label{eg}
\end{eqnarray}
satisfying the completeness relation
\begin{eqnarray}
\int d^{2}p~|p)(p|=1_{Q}~.
\label{mom_comp}
\end{eqnarray}
With the above formalism in place, we observe that
the wave-function of a ``free particle" on the NC plane is given by 
\cite{Smail},\cite{sunandan}
\begin{eqnarray}
\psi_{\vec{p}}=(p|z, \bar{z})=\frac{1}{\sqrt{2\pi\hbar^{2}}}
e^{-\frac{\theta}{4\hbar^{2}}\bar{p}p}
e^{i\sqrt{\frac{\theta}{2\hbar^{2}}}(p\bar{z}+\bar{p}z)}
\quad,\quad p=p_{x}+ip_{y}~.
\label{wavefunction}
\end{eqnarray}
The position eigenstates $|z,\bar{z})$, on the other hand, satisfy the
following completeness relation
\begin{eqnarray}
\int \frac{\theta dzd\bar{z}}{2\pi}~|z, \bar{z})\star(z, \bar{z}|=1_{Q}
\label{eg6}
\end{eqnarray}
where the Voros star product between two functions 
$f(z, \bar{z})$ and $g(z, \bar{z})$ is defined as \cite{Voros,Lizzi}
\begin{eqnarray}
f(z, \bar{z})\star g(z, \bar{z})=f(z, \bar{z})
e^{\stackrel{\leftarrow}{\partial_{\bar{z}}}
\stackrel{\rightarrow}{\partial_z}} g(z, \bar{z})~.
\label{eg7}
\end{eqnarray}
To prove this, we use (\ref{wavefunction}) and compute
\begin{eqnarray}
\int \frac{\theta dzd\bar{z}}{2\pi}
(p'|z, \bar{z})\star(z, \bar{z}|p)=
e^{-\frac{\theta}{4\hbar^{2}}(\bar{p}p+\bar{p}'p')}
e^{\frac{\theta}{2\hbar^{2}}\bar{p}p'}\delta(p-p')=(p'|p)~.
\label{eg8}
\end{eqnarray}
The completeness relation for the position eigenstates in (\ref{eg6}) 
implies that a consistent  probabilistic interpretation of finding
the particle at position $z$ can
be given iff the point-wise multiplication between the complex
conjugated wave-function and the wave-function (\ref{wavefunction}) 
is replaced by the Voros product \cite{gouba}:
\begin{eqnarray}
P(z)\propto (p|z) 
e^{\stackrel{\leftarrow}{\partial_{\bar{z}}}
\stackrel{\rightarrow}{\partial_z}}(z|p)~.
\label{interpretation}
\end{eqnarray}
The computation of the above expression shows that the probability
$P(z)$ is independent of $z$ and $p$ as expected for a free particle.
Once we have these observations and interpretations in place,
we now move on to write down the overlap of two coherent states 
$|\xi, \bar{\xi})$ and $|w, \bar{w})$ using the
completeness relation for the position eigenstates in (\ref{eg6}) 
\begin{eqnarray}
(w, \bar{w}|\xi, \bar{\xi})=\int \frac{\theta dzd\bar{z}}{2\pi}
~(w, \bar{w}|z, \bar{z})\star(z, \bar{z}|\xi, \bar{\xi})~.
\label{overlap}
\end{eqnarray}
A simple inspection shows that the following solution satisfies the above
equation
\begin{eqnarray}
(w, \bar{w}|z, \bar{z})=\frac{1}{\theta}e^{-r^2/(2\theta)}
\quad;\quad r=\sqrt{2\theta}|\omega-z|~.
\label{overlap2}
\end{eqnarray}
The Voros product, therefore, gives a specific representation of the Dirac
delta function since
\begin{eqnarray}
\lim_{\theta\rightarrow0}\frac{1}{\theta}e^{-r^2/(2\theta)}
=2\pi\delta^{(2)}(r)~.
\label{limit1}
\end{eqnarray}
Correspondingly, in three space dimensions, a similar representation for
the delta function would be
\begin{eqnarray}
\lim_{\theta\rightarrow0}\frac{1}{(4\pi\theta)^{3/2}}
e^{-r^2/(4\theta)}=\delta^{(3)}(r)~.
\label{limit2}
\end{eqnarray}
This motivates one to write down the
mass density of a static, spherically symmetric,
smeared, particle-like gravitational source in three space dimensions as 
\begin{eqnarray}
\rho_{\theta}(r)=\frac{M}{(4\pi\theta)^{3/2}}
\exp\left(-\frac{r^2}{4\theta}\right)~.
\label{massden}
\end{eqnarray}  
The above arguements clearly point out the important
role played by the Voros product in defining the mass density of the NC
Schwarzschild black hole. Indeed our analysis provides a heuristic derivation
and a possible justification for choosing (\ref{massden}) as the mass
density which is otherwise unclear in the original literature
\cite{Smail,Smailrev,Majhi1,Modaknon}.

Solving Einstein's equations with the above mass density incorporated
in the energy-momentum tensor leads to the following 
NC Schwarzschild metric \cite{Smailrev},\cite{Smail}
\begin{eqnarray}
ds^2 = -f_{\theta}(r) dt^2 + f_{\theta}^{-1}(r)dr^2 + 
r^2(d\tilde\theta^2+\sin^2\tilde\theta d\phi^2) 
\label{1.04}
\end{eqnarray}
where, 
\begin{eqnarray}
g_{tt}(r)=g^{rr}(r)=f_{\theta}(r)=\left(1-\frac{4M}{r\sqrt\pi}\gamma(\frac{3}{2},\frac{r^2}{4\theta})\right).
\label{metric_coef}
\end{eqnarray}

\noindent The event horizon of the black hole can be 
found by setting $g_{tt}(r_h)=0$ in (\ref{1.04}), which yields
\begin{eqnarray}
r_h=\frac{4M}{\sqrt\pi}\gamma(\frac{3}{2},\frac{r^2_h}{4\theta}).
\label{1.05}
\end{eqnarray}
Since this equation cannot be solved in a closed form we take 
the large radius regime ($\frac{{r_h}^2}{4\theta}>>1$) 
where we can expand the incomplete gamma function to 
solve $r_h$ by iteration. Keeping upto the leading order 
$\frac{1}{\sqrt{\theta}}e^{-{M^2}/{\theta}}$, we find
\begin{eqnarray}
r_h \simeq 2M\left(1-\frac{2M}{\sqrt{\pi\theta}}e^{{-M^2}/{\theta}}\right)~. 
\label{1.06}
\end{eqnarray}


\section{Corrected area law from quantum tunneling}

Now for a general static and spherically symmetric spacetime 
the Hawking temperature ($T_H$) is related 
to the surface gravity ($\kappa$) by the following relation \cite{Majhi1}
\begin{eqnarray}
T_H=\frac{\hbar\kappa}{2\pi} 
\label{1.061}
\end{eqnarray}
where the surface gravity of the black hole is given by
\begin{eqnarray}
\kappa = \frac{1}{2}\left(\frac{df_{\theta}}{dr}\right)_{r=r_h}.
\label{1.07}
\end{eqnarray}
Therefore the Hawking temperature for the 
noncommutative Schwarzschild black hole is found to be
\begin{eqnarray}
T_H &=& {\frac{\hbar}{4\pi}}\left[{\frac{1}{r_h}}-
{\frac{r_h^2}{4\theta^{3/2}}}\frac{e^-{\frac{{r_h}^2}{4\theta}}}
{\gamma({\frac{3}{2}},{\frac{r^2_h}{4\theta}})}\right].
\label{1.08}
\end{eqnarray}
To write the Hawking temperature in the regime 
$\frac{r^2_h}{4\theta}>>1$ as a function of $M$ 
we use (\ref{1.06}). Keeping upto the leading order in $\theta$, we get
\begin{eqnarray}
T_{H}\simeq\frac{\hbar}{8{\pi}M}
\left[1-\frac{4M^3}{{\sqrt\pi} {\theta}^{3/2}}{e^{-M^2/\theta}}\right].
\label{1.10}
\end{eqnarray}
We shall now use the first law of black hole thermodynamics 
to calculate the Bekenstein-Hawking entropy. 
The first law of black hole thermodynamics is given by  
\begin{eqnarray}
dS=\frac{dM}{T_H}~.
\label{1.1}
\end{eqnarray}
Hence the Bekenstein-Hawking entropy upto  
leading order in $\theta$ is found to be 
\begin{eqnarray}
S=\int{\frac{dM}{T_H}}=\frac{1}{\hbar}
(4\pi M^2-16M^3\sqrt{\frac{\pi}{\theta}}
e^{-\frac{M^2}{\theta}})+\mathcal{O}(\sqrt{\theta}e^{-\frac{M^2}{\theta}}).
\label{1.11}
\end{eqnarray}
The same expression of Bekenstein-Hawking entropy was 
found earlier in \cite{Majhi1} by the tunneling method. 
In order to express the entropy in terms of the 
noncommutative horizon area ($A_{\theta}$), we use (\ref{1.06}) to obtain
\begin{eqnarray}
A_{\theta} = 4\pi r^2_h=16\pi M^2-64\sqrt{\frac{\pi}{\theta}}
M^3e^{-\frac{M^{2}}{\theta}}.
\label{1.12}
\end{eqnarray}
Comparing equations (\ref{1.11}) and (\ref{1.12}), 
we find that at the leading order in $\theta$, 
the noncommutative black hole entropy satisfies the area law
\begin{eqnarray}
S=S_{\textrm{BH}}=\frac{A_{\theta}}{4\hbar}~.
\label{1.13}
\end{eqnarray}
This is functionally identical to the Benkenstein-Hawking 
area law in the commutative space.
     
\noindent Hence we have analytically observed that in the 
regime $\frac{r^2_h}{4\theta}>>1$, the noncommutative version 
of the semiclassical Bekenstein-Hawking area law holds 
upto leading order in $\theta$. This motivates us to investigate 
the corrections to the semiclassical area law upto leading order in $\theta$.

To do so, we first compute the corrected 
Hawking temperature $\tilde{T}_{H}$. 
For that we use the tunneling method by 
going beyond the semiclassical approximation 
\cite{Majhibeyond}. Considering the massless scalar particle tunneling, the Klein-Gordon equation under the background metric (\ref{1.04}) is given by
\begin{equation}
-\frac{\hbar^2}{\sqrt{-g}}{\partial_\mu[g^{\mu\nu}\sqrt{-g}\partial_{\nu}]\Phi}=0.
\label{KG}
\end{equation}   
It is worthwhile to point out that since noncommutativity is coming here through the matter sector ($\Phi$), the form of the Klein-Gordon equation does not change with respect to the spacetime coordinates. For simplicity we restrict ourselves to the radial trajectory so that only the $r-t$ sector of the metric (\ref{1.04}) is necessary. Since equation (\ref{KG}) cannot be solved exactly, we proceed by choosing a standard WKB ansatz for $\Phi$ as
\begin{eqnarray} 
\Phi(r,t)=\exp[-\frac{i}{\hbar}{{\cal S}(r,t)}], 
\label{ansatz}
\end{eqnarray}  
where, 
\begin{eqnarray}
{\cal S}(r,t)={\cal S}_0(r,t)+\displaystyle
\sum_{i=1}^{\infty}\hbar^{i}{\cal S}_{i}(r,t).
\label{action}
\end{eqnarray}
Now substituting (\ref{action}) in (\ref{KG}) and equating coefficients of different powers in $\hbar$ to zero one obtains a set of partial differential equations 
\cite{Majhibeyond}-\cite{Tao}, \cite{Modakex}. They can be simplified to find the solution for $n-th$ order in $\hbar$, given by the first order partial differential equation
\begin{eqnarray}
\frac{\partial S_n}{\partial t}=\pm f_{\theta}(r)
\frac{\partial S_n}{\partial r},
\label{nth}
\end{eqnarray}
where ($n=~0,~i;~i= 1,2,...)$. For the lowest order in $\hbar$ ($n=0$) we are left with a semiclassical Hamilton-Jacobi type equation 
\begin{eqnarray}
\frac{\partial S_0}{\partial t}=\pm f_{\theta}(r)
\frac{\partial S_0}{\partial r}.
\label{0th}
\end{eqnarray}
Now we can choose the functional form of the semiclassical action ${\cal S}_0(r,t)$ by looking at the symmetry of the background metric (\ref{1.04}) as
\begin{eqnarray}
{\cal S}_{0}(r,t)=\omega t+{\cal S}_{0}(r).
\label{actionform}
\end{eqnarray} 
Here $\omega$ is the conserved quantity corresponding to the time translation Killing vector field in the background metric (\ref{1.04}) and is represented by the Komar energy integral
\begin{eqnarray}
\omega=\frac{1}{4\pi}\int_{\partial\Sigma}d^{2}x~\sqrt{p^{(2)}}~n^{\mu}\sigma^{\nu}\nabla_{\mu}K_{\nu}~.
\label{komarint}
\end{eqnarray}
This is defined on the boundary ($\partial{\Sigma}$) of a spacelike hypersurface $\Sigma$ and $p_{ij}$ is the induced metric on $\partial{\Sigma}$, while $p^{(2)}={\textrm{det}}~p_{ij}$. Unit normal vectors $n^{\mu}$ and $\sigma^{\nu}$ are associated with $\Sigma$ and $\partial{\Sigma}$ respectively, whereas, $K_{\nu}$ is the timelike Killing vector. When observed at infinity this is the energy of the spacetime and matches with the commutative mass ($M$) of the black hole. But unlike the standard Schwarzschild black hole, in this case the effective energy experienced by a particle at finite distance is not the same as experienced at infinity. Because of the noncommutative effects it is modified at finite distances. It will be discussed more elaborately later on (in section 4) in order to find the leading correction to the semiclassical entropy. For now we proceed with the calculation of corrected temperature and entropy of the black hole (\ref{1.04}). Putting (\ref{actionform}) in (\ref{0th}) we have
\begin{eqnarray}
{\cal S}_{0}(r)=\pm\omega\int_C\frac{dr}{f_{\theta}(r)},
\label{s0}
\end{eqnarray}
where the $+ (-)$ sign stands for the ingoing (outgoing) particle. The contour $C$ is chosen such that it starts from just behind the event horizon to the outer region, left to right, in the lower half of the complex plane, avoiding the singularity at the event horizon. Using (\ref{actionform}) and (\ref{s0}) we finally find the semiclassical action as
\begin{eqnarray}
{\cal S}_{0}(r,t)=\omega(t\pm\int_C\frac{dr}{f_{\theta}(r)}).
\label{action2}
\end{eqnarray} 
Since for all $n$$, {\cal S}_n(r,t)$-s satisfy the similar type of differential equations, the solutions for any ${\cal S}_i(r,t)$ can differ from ${\cal S}_{0}(r,t)$ only by a proportionality constant. The most general solution for the scalar particle action is then given by
\begin{eqnarray}
{\cal S}(r,t) = (1+\displaystyle\sum_{i=1}^{\infty}\gamma_i \hbar^i) {\cal S}_0(r,t),
\label{fullaction1}
\end{eqnarray}
where $\gamma_i$-s are the proportionality constants and have the dimensions of $\hbar^{-i}$. Since in (3+1) dimensions in the unit $c=G=\kappa_{B}=1$, Planck length ($l_p$) and Planck mass ($m_p$) {\footnote{$l_p^2=\frac{\hbar G}{c^3},m_p^2=\frac{\hbar c}{G}$}} are proportional to $\sqrt\hbar$, we can readily express (\ref{fullaction1}) as
\begin{eqnarray}
{\cal S}(r,t) &=& (1+\displaystyle\sum_{i=1}^{\infty}{\frac{\tilde\beta_i \hbar^i}{(Mr_h)^i})} {\cal S}_0(r,t) \nonumber\\
              &=& \omega(1+\displaystyle\sum_{i=1}^{\infty}{\frac{\tilde\beta_i \hbar^i}{(Mr_h)^i})}(t\pm\int_C\frac{dr}{f_{\theta}(r)}),
\label{fullaction}
\end{eqnarray}     
where $\tilde\beta_i$-s are dimensionless constants which can be related to the trace anomaly of the stress tensor of the scalar field. The ingoing ($\Phi_{in}$) and outgoing ($\Phi_{out}$) scalar field modes can now be found by substituting ${\cal S}(r,t)$ from (\ref{fullaction}) into (\ref{ansatz}) with proper choice of sign. As a result the ingoing and outgoing probabilities are given by
\begin{eqnarray}
P_{in} = |\Phi_{in}|^2 &=& {\exp}\Big[\frac{2}{\hbar}(1+\sum_i\tilde\beta_i\frac{\hbar^i}{(M r_h)^i})\Big(\omega{\textrm{Im}}~t +\omega{\textrm{Im}}\int_C\frac{dr}{f_{\theta}(r)}\Big)\Big]\label{trace1}\\
P_{out} = |\Phi_{out}|^2 &=& {\exp}\Big[\frac{2}{\hbar}(1+\sum_i\tilde\beta_i\frac{\hbar^i}{(M r_h)^i})\Big(\omega{\textrm{Im}}~t -\omega{\textrm{Im}}\int_C\frac{dr}{f_{\theta}(r)}\Big)\Big].\label{trace2}
\end{eqnarray}
respectively. Since the incoming mode can always get inside the event horizon, one has $P_{in}=1$, which gives $\textrm{Im}~t =-{\textrm{Im}}\int_C\frac{dr}{f_{\theta}(r)}$. Substituting this in (\ref{trace2}) we get
\begin{eqnarray}
P_{out} = {\exp}\Big[-\frac{4\omega}{\hbar}(1+\sum_i\tilde\beta_i\frac{\hbar^i}{(M r_h)^i}){\textrm{Im}}\int_C\frac{dr}{f_{\theta}(r)}\Big].
\label{pout} 
\end{eqnarray}
Now the corrected Hawking temperature for the noncommutative Schwarzschild black hole can be identified by using the principle of detailed balance \cite{Paddy} which relates the ingoing and outgoing probabilities as $P_{out}=\exp\big({-\frac{\omega}{\tilde T_{H}}}\big)~P_{in}$. Taking $P_{in}$ to be unity and $P_{out}$ from (\ref{pout}) one can calculate the corrected Hawking temperature as
\begin{eqnarray}
\tilde{T}_{H}=\frac{\hbar}{4}\left[1+\sum_{i}\frac{{\tilde\beta}_{i}\hbar^{i}}{(Mr_{h})^{i}}\right]^{-1}~\left({\textrm{Im}}\int_C\frac{dr}{f_{\theta}(r)}\right)^{-1}.
\label{corrtemp}
\end{eqnarray}
Since the tunneling phenomena takes place in the close vicinity of the event horizon, we can express $f_{\theta}(r)=f_{\theta}(r_h)+(r-r_h)f'_{\theta}(r_h)+{\cal O}(r-r_h)^2=(r-r_h)f'_{\theta}(r_h)+{\cal O}(r-r_h)^2$. Substituting this in (\ref{corrtemp}) and performing the contour integral we finally get the corrected Hawking temperature as
\begin{eqnarray}
\tilde{T}_{H}=T_{H}\left[1+\sum_{i}\frac{\tilde{\beta}_{i}\hbar^{i}}
{(Mr_{h})^{i}}\right]^{-1}~.
\label{corr_temp}
\end{eqnarray}
Hence, once again applying the first law of black hole thermodynamics with this corrected Hawking temperature, we obtain the following expression for the corrected entropy/area law :
\begin{eqnarray}
S_{bh}  &=& \frac{A_{\theta}}{4\hbar}+2\pi\tilde{\beta}_{1}\ln A_{\theta} - \frac{16\pi^{2}\tilde{\beta}_{2}\hbar^2}{A_{\theta}}+\mathcal{O}(\sqrt{\theta}e^{-\frac{M^2}{\theta}})~\nonumber\\        &=& S_{BH}+2\pi\tilde{\beta}_{1}\ln S_{BH}-\frac{4\pi^{2}\tilde{\beta}_{2}\hbar}{S_{BH}}+\mathcal{O}(\sqrt{\theta}e^{-\frac{M^2}{\theta}}),
\label{corr_entr}
\end{eqnarray}
where $A_{\theta}$ and $S_{BH}$ are defined in (\ref{1.12}) and (\ref{1.13}) respectively. This expression is functionally identical to the corrected entropy/area law for the standard Schwarzschild black hole \cite{Majhitrace,Modakex}. However there is an important difference. This expression of corrected entropy has both noncommutative and quantum corrections. Although here we have restricted ourselves only to the leading order correction due to the NC parameter ($\theta$), one can try to include all order $\theta$ corrections. 
This is technically more involved and we shall not 
address this issue in this paper. 
Now we move on to the next section to compute 
the coefficient $\tilde{\beta}_{1}$ in the above expression.

\section{Calculation of the coefficient $\tilde\beta_1$}
By making an infinitesimal scale transformations to the metric coefficients in (\ref{1.04}),  the coefficient $\tilde{\beta}_{1}$ can be related to the trace anomaly in the following way \cite{Modakex}:
\begin{eqnarray}
\tilde{\beta}_{1} &=& -\frac{M}{4\pi\omega}{\textrm{Im}}\int d^{4}x~\sqrt{-g}~\langle{T^{\mu}}_{\mu}\rangle^{(1)}~\nonumber\\
                  &=& -\frac{M}{4\pi\omega}{\textrm{Im}}\int_{r_h}^{\infty}\int_{0}^{i\beta}\int_{0}^{\pi}\int_{0}^{2\pi} r^2 \sin{\tilde\theta}\langle{T^{\mu}}_{\mu}\rangle^{(1)} drdtd\tilde\theta d\phi       
\label{coeff_1}
\end{eqnarray}
Here, $\langle{T^{\mu}}_{\mu}\rangle^{(1)}$ is the trace anomaly calculated for the first loop expansion and $\omega$ is given by the Komar energy integral (\ref{komarint}) evaluated near the event horizon.
The one loop trace anomaly of the stress tensor for the scalar fields moving in the background of a (3+1) dimensional curved manifold is given by \cite{Hawking3,Dewitt}
\begin{eqnarray}
\langle{T^{\mu}}_{\mu}\rangle^{(1)}=\frac{1}{2880\pi^{2}}
\left(R_{abcd}R^{abcd}-R_{ab}R^{ab}+\nabla_{a}\nabla^{a}R\right)~.
\label{trace_anomaly}
\end{eqnarray}
For the metric (\ref{1.04}), the invariant scalars are explicitly found as 
\begin{eqnarray}
R_{abcd}R^{abcd} &=& \frac{48M^2}{r^6}+\frac{M^2e^{-(r^2/2\theta)}}
{4\pi r^6\theta^5}\times\nonumber\\
&& \left[r^{10}+16\alpha_1+32\theta^3e^{(r^2/4\theta)}\Gamma(\frac{3}{2},\frac{r^2}{4\theta})\alpha_2+768e^{(r^2/4\theta)}\Gamma(\frac{3}{2},\frac{r^2}{4\theta})\right]
\label{Rabcd}\\
&&{\textrm{where,}}~~~ \alpha_1=(r^6\theta^2-\sqrt\frac{\pi}{\theta}r^5\theta^3e^{(r^2/4\theta)}-4\sqrt\frac{\pi}{\theta}r^3\theta^4e^{(r^2/4\theta)})\nonumber\\
&& \alpha_2=\left(\frac{r^5}{\sqrt\theta}-24\sqrt\pi\theta^2e^{(r^2/4\theta)}+4\theta^2(\frac{r^2}{\theta})^{3/2}\right)
\nonumber
\end{eqnarray}
\begin{eqnarray}
R_{ab}R^{ab}=\frac{M^2e^{-(r^2/2\theta)}(r^4-8r^2\theta+32\theta^2)}{8\pi\theta^5}\\
R=-\frac{Me^{-(r^2/4\theta)}(r^2-8\alpha)}{2\sqrt\pi\theta^{5/2}}
\label{Rab} 
\end{eqnarray}
Note that in the commutative limit ($\theta\rightarrow 0$) the above results match with the known results of the standard vacuum Schwarzschild spacetime metric, for which $R_{abcd}R^{abcd}=\frac{48M^2}{r^6},~~R_{ab}R^{ab}=0,~~R=0$. 
To find the trace anomaly (\ref{trace_anomaly}), we now evaluate
\begin{eqnarray}
\nabla_a\nabla^aR &=& -\frac{Me^{-r^2/2\theta}}{8\pi\theta^5(r^2/\theta)^{1/2}}\left[2M(r^2-12\theta)(r^2/\theta)^{3/2}\theta+\sqrt\pi e^{r^2/4\theta}\alpha_3+4Me^{(r^2/4\theta)}\Gamma(3/2,r^2/4\theta)\alpha_4\right]~~~~~\label{nabla}\\
&& {\textrm{where,}}~~~\alpha_3 = \left(r^5-22r^3\theta+72r\theta^2-2M(r^4-20r^2\theta+48\theta^2)\right)\nonumber\\
&& \alpha_4 = (r^4-20r^2\theta+48\theta^2).\nonumber 
\end{eqnarray}
Exploiting all these results, the trace anomaly is calculated 
from (\ref{trace_anomaly}), upto the leading order in 
${\cal O}(e^{-\frac{r^2}{4\theta}})$, as 
\begin{eqnarray}
\langle{T^{\mu}}_{\mu}\rangle^{(1)}=\frac{1}{2880\pi^{2}}
\left[\left(\frac{48M^{2}}{r^{6}}-\frac{4M^2}{\sqrt{\pi}\theta^{5/2}}
\frac{e^{-r^{2}/(4\theta)}}{r}\right)
-\frac{Me^{-r^{2}/(4\theta)}}{8\sqrt{\pi}\theta^{9/2}}[r^4-2Mr^3-22r^{2}
\theta+40Mr\theta]\right]\nonumber\\
+\mathcal{O}(e^{-r^{2}/(2\theta)})~.
\label{trace_anomaly_1}
\end{eqnarray}
Substituting this in (\ref{coeff_1}) and performing the integral yields
\begin{eqnarray}
\tilde{\beta}_{1}=\frac{M}{180\pi\omega}
\left[1+\frac{2M}{\sqrt{\pi\theta}}(1-\frac{2M^2}{\theta})
e^{-M^{2}/(\theta)}\right]+\mathcal{O}(\sqrt{\theta}e^{-M^{2}/(\theta)})~.
\label{coeff_1a}
\end{eqnarray}
To compute $\omega$, we calculate the Komar energy 
integral (\ref{komarint}). For the spacetime metric (\ref{1.04}) one has
the following expressions for the Killing vectors ($K^\mu$), its inverse
($K_\nu$) and the unit normal vectors ($n^\mu$, $\sigma^\nu$)
\begin{eqnarray}
&& K^{\mu} = (1,0,0,0),~~~K_{\nu} = -f_{\theta}(1,0,0,0)\\
&& n^{\mu} = f_{\theta}^{-1/2}(1,0,0,0)\\
&& \sigma^{\nu} = f_{\theta}^{1/2}(0,1,0,0)\\
&& \sqrt{p^{(2)}} = r^2\sin\theta.
\label{kom}
\end{eqnarray}
Using these, eq.(\ref{komarint}) is simplified as
\begin{eqnarray}
\omega=\frac{1}{8\pi}\int_{\tilde\theta=0}^{\pi}\int_{\phi=0}^{2\pi}
r^2\sin\tilde\theta~(\partial_{r}f_{\theta})~d\tilde\theta d\phi
\label{komarsimp}
\end{eqnarray}
Finally, integrating over the angular variables, we get 
\begin{eqnarray}
\omega=M\left[1-\frac{r}{\sqrt{\pi\theta}}(1+\frac{r^2}{2\theta})
e^{-r^{2}/(4\theta)}\right]+\mathcal{O}(\sqrt{\theta}e^{-r^{2}/(4\theta)})~.
\label{komar1a}
\end{eqnarray}
Note that at spatial infinity ($r\rightarrow\infty$), 
the Komar energy ($\omega$) is nothing but the 
commutative mass ($M$) of the spacetime as expected. 
However for finite distance the effective energy involves NC corrections. Also, for the commutative limit ($\theta\rightarrow 0$), $\omega= M$ holds at any radial distance outside the event horizon which is a well known fact for the standard Schwarzschild black hole. Near the event horizon (\ref{1.06}), the above expression for $\omega$ simplifies to
\begin{eqnarray}
\omega=M\left[1-\frac{2M}{\sqrt{\pi\theta}}(1+\frac{2M^2}{\theta})
e^{-M^{2}/\theta}\right]+\mathcal{O}(\sqrt{\theta}e^{-M^{2}/(\theta)})~.
\label{komar1b}
\end{eqnarray}
Substituting this in the expression for $\tilde{\beta}_{1}$ 
in (\ref{coeff_1a}), we obtain :
\begin{eqnarray}
\tilde{\beta}_{1}=\frac{1}{180\pi}
\left[1+\frac{4M}{\sqrt{\pi\theta}}e^{-M^{2}/\theta}\right]+\mathcal{O}(\sqrt{\theta}e^{-M^{2}/\theta})~.
\label{coeff_1b}
\end{eqnarray}
Exploiting (\ref{corr_entr}) and (\ref{coeff_1b}), 
we find the cherished result for the corrected entropy/area law (upto leading order correction) for the NC Schwarzschild black hole
\begin{eqnarray}
S_{bh}  &=& \frac{A_{\theta}}{4\hbar}+\frac{1}{90}(1+\frac{4M}{\sqrt{\pi\theta}}e^{-M^2/\theta})\ln\frac{A_{\theta}}{\hbar} + \mathcal{O}(\sqrt{\theta}e^{-\frac{M^2}{\theta}})
\nonumber\\        
        &=& S_{BH}+\frac{1}{90}(1+\frac{4M}{\sqrt{\pi\theta}}e^{-M^2/\theta})\ln S_{BH}+\mathcal{O}(\sqrt{\theta}e^{-\frac{M^2}{\theta}})
\label{corr_entr2}
\end{eqnarray}
This is the general expression for the entropy of NC 
Schwarzschild black hole where both the NC and quantum 
effects have been taken into account. 
The first term in this expression is the semiclassical 
entropy and the next term is the leading correction. 
It is logarithmic in nature. The coefficient of the logarithmic correction is different from the standard Schwarzschild black hole \cite{Majhitrace, Modakex} due to the presence of noncommutative parameter ($\theta$). In the commutative limit $\theta\rightarrow 0$, the expression for the corrected entropy exactly matches with the standard Schwarzschild case where the coefficient of the leading correction is $\frac{1}{90}$, obtained in the path integral \cite{Hawking3}, euclidean \cite{fur} and tunneling \cite{Majhitrace},\cite{Modakex} formalisms.

\section{Conclusions}
We now conclude by making the following comments. 
In this paper we have shown the importance of the 
Voros star product in writing down the mass density 
of a noncommutative Schwarzschild black hole. 
To point out the role played by the Voros product, 
we have first taken recourse to a rigorous formulation 
of NC quantum mechanics \cite{gouba},\cite{sunandan}. 
It has been observed that the inner product of the 
coherent states (used in the construction of the 
wave-function of a ``free point particle"), can be 
computed by using a deformed completeness relation 
(involving the Voros product) among the coherent states. 
This inner product is then used to write down the 
mass density by making a dimensional lift 
from two to three space dimensions. 
A consistent probabilistic interpretation of the 
wave-function of the ``free point particle" can also 
be given only when the Voros product is incorporated. 
The Gaussian distribution of the mass density indeed 
follows from this Voros type interpretation and 
naturally includes the effect of the NC parameter 
$\theta$ and agrees with the structure given in the 
literature \cite{Modaknon}-\cite{Smail}, \cite{lopez}.

Another part of the paper dealt with the entropy/area law 
corrections of the NC Schwarzschild black hole. 
A general result for NC Schwarzschild black hole 
entropy/area law was found, taking both quantum 
and NC effects into account. For this we first 
used the tunneling method by going beyond the 
semiclassical approximation and calculated the 
corrected Hawking temperature. 
This result involved the (NC) semiclassical Hawking temperature 
at the lowest order and corrections at higher orders. 
Using this modified temperature and the 
first law of black hole thermodynamics we then calculated 
the corrected entropy. The (NC) semiclassical 
Bekenstein-Hawking value was reproduced at the 
lowest order and higher order corrections contained 
logarithmic and inverse powers of horizon area. 
The coefficient of the leading (logarithmic) 
term was fixed by using the trace anomaly of the scalar 
field stress tensor. The trace anomaly and the 
Komar energy integral for NC Schwarzschild 
metric were explicitly calculated to determine 
this coefficient. The value of the 
coefficient was found to have NC correction. 
We also show that the commutative limit of 
the corrected entropy/area law of the NC Schawrzschild black 
hole matches with the standard result 
for the Schwarzschild black hole \cite{Hawking3,fur,Majhitrace,Modakex}.

\section*{Acknowledgements} One of the authors (SKM) thanks the Council
of Scientific and Industrial Research (CSIR), Government of India,
for financial support.

\end{document}